\documentclass[twocolumn,preprintnumbers,amsmath,amssymb,aps]{revtex4}
\usepackage{graphicx}
\usepackage{dcolumn}
\usepackage{bm}
\usepackage{color}
\begin{document}
\preprint{}
\title{From nuclear matter to Neutron Stars}
\author{T. K. Jha {\footnote {email: tkjha@prl.res.in}}}
\affiliation 
{
Theoretical Physics Division, Physical Research Laboratory, Ahmedabad, India - 380 009
}
\date{\today}
\begin{abstract}

Neutron stars are the most dense objects in the observable Universe and conventionally one uses nuclear theory to obtain the equation of state (EOS) of dense hadronic matter and the global properties of these stars. In this work, we review various aspects of nuclear matter within an effective Chiral model and interlink fundamental quantities both from nuclear saturation as well as vacuum properties and correlate it with the star properties.

\end{abstract}
\maketitle
\section{Introduction}

\textit{\textbf{Aspects of Nuclear Matter:-}}
Nuclear matter is a hypothetical, infinitely large system of nucleons ($N=Z$) with coulomb interaction turned off. The matter inside the compact stars, the early universe, a nucleon gas, 
or quark-gluon plasma ({\bf QGP}) are some of the types of nuclear matter. The only physical parameters that remains to characterize such a system is the bulk binding energy and the saturation density, letting the total number of nucleons go to infinity.

\begin{figure}[ht]
\vskip -0.2in
\begin{center}
\includegraphics[width=7cm,angle=0]{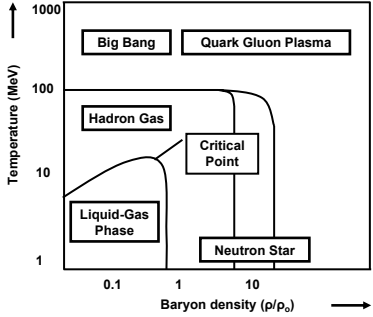}
\vskip -0.2in
\caption{Theoretical Phase diagram for nuclear matter.}
\end{center}
\end{figure}

Fig. 1 shows the conventional picture of different aspects and phase of nuclear matter with the variation of baryon density ($\rho_B$) and temperature (T). At low temperature and high density, roughly of the order of ten times normal nuclear density, we approach the matter configuration which may be similar to that present in the core of neutron stars. The discovery of neutron stars in the form of pulsars has been a major stimulus to dense matter studies, which serves as an ideal laboratory to study matter at extreme conditions. They are the most compact objects observed till date, confined to a radius of nearly 10 km with the density ranging from $(3~-~10)~\rho_0$ ($\rho_0$ is the normal nuclear matter density) in the core of these stars. Bound by gravity, the evolution and constitution of these stars represents a beautiful amalgamation of all the known forces in nature. 

Theoretically, the framework of Quantum Hadrodynamics \cite{wal74,serot86} present an elegant and consistent treatment of finite nuclei as well as infinite nuclear matter, which seem to provide solution to the so called ``{\it the Coester band}" problem \cite{co01}. A realistic nuclear equation of state must satisfy certain minimum criteria quantified as the {\it ``nuclear saturation properties''}, which are the physical constants of nature. In this context, the two most important quantities which play vital role and are known to have substantial impact on the EOS are the nucleon effective mass and the nuclear incompressibility \cite{compact}, which are not very well determined and posses large uncertainty. Nucleon effective mass or the medium mass modification of nucleon in nuclear medium is a consequence of the Dirac field and forms an essential element for the success of relativistic phenomenology. On the other hand, the nuclear incompressibility derived from nuclear measurements and astrophysical observations exhibit a broad range of values $K = (180 - 800)$ MeV \cite{glen88}. Further the non-relativistic and the relativistic models fails to agree to a common consensus. The non-relativistic calculations predict the compression moduli in the range $K=(210-240) MeV$ \cite{k2}, whereas, relativistic calculations predicts it in the range $(200-300)~MeV$ \cite{nl3,k6}. Apart from that we are inevitably marred by the uncertainty in the determination of mass of the scalar meson ($\sigma$-meson). The attractive force resulting from the scalar sector is responsible for the intermediate range attraction which, along with the repulsive vector forces provides the saturation mechanism for nuclear matter \cite{serot86}. The estimate from the Particle Data Group quotes the mass of this scalar meson `$f_0 (600)$' or $\sigma-$meson in the range $(400 - 1200)$ MeV \cite{pdg}. A recent estimate however, for sigma meson mass is found to be $513 \pm 32$ MeV \cite{mura02}. Therefore in order to address these issues, one needs a model that has the desired attributes of the relativistic framework and which can be successfully applied to various nuclear force problem both in the vicinity of $\rho_0$ as well as at higher densities.

\vskip 0.1in
\textit{\textbf{Neutron Star Structure:-}} Matter at extremes opens up numerous possibilities, which may be present in neutron star core as shown in Fig. 2. Possible scenarios such as the transition from nuclear and hadronic matter, to exotic states involving pionic and  Kaonic Bose-Einstein condensation, to bulk quark matter and quark matter in droplets, including super-conducting states, as well as strange quark matter, have been proposed \cite{compact,weber,QCD}. Hyperons can form in the neutron star cores, when the nucleon chemical potential are large enough to compensate the mass difference between the nucleon and the hyperon. Also, in dense matter, the energy of the pion ($\pi$ meson) and the kaon ($K$) is modified by inter-particle interactions and if it becomes sufficiently low, they will form the condensate, i.e., being bosons, they will accumulate the same ground state \cite{condense}. The presence of the condensates is also known to enhance the neutrino cooling \cite{cool}, in addition to the nucleon direct URCA (Unrecordable Cooling Agents) process \cite{urca}. It is also widely believed that the attractive interactions between the quarks will lead to pairing and color superconductivity. For three massless flavors, the condensation pattern that minimizes the free energy is known as the color-flavor-locked (CFL) scheme \cite{cfl}. A detailed description of these novel phases can be found in Ref. \cite{sanjay, strange1}.

\begin{figure}\centering
\includegraphics[width=5cm,angle=0]{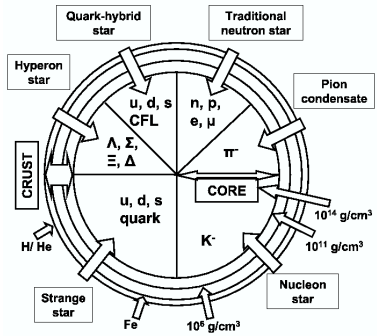}
\caption{Possible phases structure in neutron star core.}
\end{figure}

\vskip 0.1in
\textit{\textbf{Observations from Neutron star:-}}
Observations from neutron stars can lead to an understanding of the state of their interiors and the key unknowns such as the stars maximum mass and radius. Apart from that, the other key observables are the gravitational redshift, the central density of the star, the moments of inertia and the pulsar timings. These informations can be inferred from the photons, ranging from radio waves to X-rays, and also those involving neutrinos and gravity waves. The largest precisely known neutron star mass is only $1.44~ M_{\odot}$, but in case of binaries with white dwarf companions, the mass might be somewhere in the vicinity of 2 $M_{\odot}$ or even larger. The radius of the star is found to be sensitive to the properties of the nuclear matter near the saturation density ($\rho_0$), such as the density dependence of the nuclear symmetry energy \cite{lat01}. However, the radius measurements are not that precisely known unlike the mass observations, but the upper limits of the radius values are inferred from the thermal emission of cooling neutron stars, the gravitational redshift measurements, or from the crustal properties such as the pulsar glitches, star quakes and cooling timescales. The vital information from these observations can be used to constrain the nuclear equation of state at high densities.

\begin{figure}
\begin{center}
\includegraphics[width=7cm,angle=0]{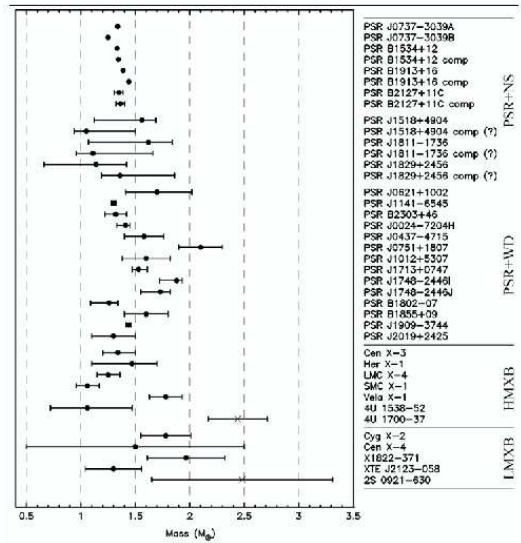}
\caption {Observed mass of neutron stars in binaries. The figure is taken from Ref \cite{sanjay}. ($PSR+NS$) is the binary of a pulsar and a neutron star and ($PSR+WD$) is the binary of a pulsar with a white dwarf companion. HMXBs' are the high-mass X-ray binaries and `LMXBs' are the low-mass X-ray binaries. The two objects plotted as crosses, the HMXB 4U $1700-37$ and the LMXB 2S $0921-630$ are blackhole candidates, but may be neutron stars. Error bars are $1\sigma$ errors.}
\end{center}
\end{figure}

The measured mass of 40 neutron stars in binary systems are summarized in Fig. 3. In case of `$PSR J0737-3039$', for the A and B system, both neutron stars are detected as radio pulsars, whereas in the three systems with the companion marked by `?' may be a white dwarf. In a high-mass X-ray binary (HMXB), the companion is a massive star $M > 10M_{\odot}$, whereas in a low-mass X-ray binary (LMXB), the companion's mass is below $1M_{\odot}$. The most accurately measured masses are from timing observations of the radio binary pulsars such as in case of binary pulsar PSR 1913+16, in which the masses are $M=1.3867 \pm 0.0002 M_{\odot}$ and $M=1.4414 \pm 0.0002 M_{\odot}$, respectively. It is found that the mass determinations in binaries with white dwarf companions shows a broader range of neutron star masses than binary neutron star pulsars. The neutron star, PSR J0751+1807 with mass 2.1 $M_{\odot}$, is about 4$\sigma$ from the canonical value of 1.4 $M_{\odot}$. The system of $Vela X-1$ however, has lower mass limit (1.6 to 1.7$M_{\odot}$).

{\it Observed radii:} From the study of the measured flux of the nearest neutron star $RX J1856.5-3754$, with the Hubble space telescope, Kaplan et.al., estimated the stars radius
\cite{ob5}. The analysis puts the radiation radius of the star $R_{\infty}$ = $\frac{R}{\sqrt{1-2GM/R}} \simeq 15$km corresponding to $R\simeq12$ km for $M=1.4M_{\odot}$. Recently, a few surveys from the non-accreting X-ray binaries, the so called the Quiescent stars 
$CXOU132619.7-472910.8$ in $NGC 5139$ gives $R_{\infty}=14.3 \pm 2.1$km (90\% CL) \cite{ob6}. In case of the neutron star $4U1636-53$, the compactness parameter `$M/R$' $<0.16$ implies a radius $R>12-13$km for the star with $M=1.4M_{\odot}$.

Some other observables from compact stars are the gravitational redshift measurement of the star, the moment of inertia or the estimated central density of the star. Overall, there are large uncertainties in these estimates. Further stringent conditions can be imposed on the EOS of dense matter from the laboratory estimates of the heavy-ion data, such as constraints on nuclear incompressibility, symmetry energy or nucleon effective mass.

\section{Effective chiral Model}

Chiral models \cite{ch01,ch02} have been developed and were applied to nuclear matter studies. We start with an effective chiral Lagrangian and calculate the density effects in nuclear matter in the mean field approach. To have a realistic description of the dense neutron 
star matter, we consider the effective Lagrangian of the chiral model generalized 
to include the lowest lying octet of baryons ($n,p,\Lambda^{0},\Sigma^{-,0,+},\Xi^{-,0}$) 
interacting through the exchange of the pseudo-scalar meson $\pi$, the scalar 
meson $\sigma$, the vector meson $\omega$ and the iso-vector $\rho-$meson, 
and is given by \cite{tkj06}:

\begin{eqnarray}
\label{lag}
{\cal L}&=& \bar\psi_B~\left[ \big(i\gamma_\mu\partial^\mu
         - g_{\omega B}\gamma_\mu\omega^\mu
         - \frac{1}{2}g_{\rho B}{\vec \rho}_\mu\cdot{\vec \tau}
            \gamma^\mu\big ) \right]~ \psi_B \nonumber \\
         &-& \bar\psi_B~\left[g_{\sigma B~}~\big(\sigma + i\gamma_5
             \vec \tau\cdot\vec \pi \big)\right]~ \psi_B
        + \frac{1}{2}\big(\partial_\mu\vec \pi\cdot\partial^\mu\vec\pi
\nonumber \\
&&
        + \partial_{\mu} \sigma \partial^{\mu} \sigma\big)
        - \frac{\lambda}{4}\big(x^2 - x^2_0\big)^2
        - \frac{\lambda B}{6}\big(x^2 - x^2_0\big)^3 \nonumber \\
        &-& \frac{\lambda C}{8}\big(x^2 - x^2_0\big)^4
      - \frac{1}{4} F_{\mu\nu} F_{\mu\nu}
        + \frac{1}{2}{g_{\omega B}}^{2}x^2 \omega_{\mu}\omega^{\mu} \nonumber \\
        &-& \frac {1}{4}{\vec R}_{\mu\nu}\cdot{\vec R}^{\mu\nu}
        + \frac{1}{2}m^2_{\rho}{\vec \rho}_{\mu}\cdot{\vec \rho}^{\mu}\ .
\end{eqnarray}

The above Lagrangian represents the interaction of baryons $\Psi_B$ with the aforesaid mesons. We have the kinetic and the non-linear terms in the pseudoscalar-isovector pion field `$\vec \pi$', the scalar field `$\sigma$', and with $x^2= {\vec \pi}^2+\sigma^{2}$. Finally in the last two lines, we have the field strength and the mass term for the vector field `$\omega$' and the iso-vector field `$\vec \rho$' meson. The terms in eqn. (1) with the subscript $'B'$ should be interpreted as sum over the states of all baryonic octets. In this paper we shall be concerned only with the normal non-pion condensed state of matter, so we take $<\vec \pi>=0$.
The interaction of the scalar and the pseudoscalar mesons with the vector boson generates a dynamical mass for the vector bosons through spontaneous breaking of the chiral symmetry with scalar field getting the vacuum expectation value $x_0$. Then the masses of the baryons, the scalar and the vector mesons, are respectively given by 

\begin{eqnarray}
m_B = g_{\sigma B} x_0,~~ m_{\sigma} = \sqrt{2\lambda} x_0,~~
m_{\omega} = g_{\omega B} x_0\ .
\end{eqnarray}

Using the meson field equations for $\omega$ and $\sigma$-meson along with the corresponding energy density and pressure expression for symmetric nuclear matter, the parameters of the present model are then evaluated. Here we fix the saturation density ($\rho_0 = 0.153~fm^{-3}$) and for a desired value of nucleon effective mass ($Y~=~m^{\star}/m~=~(0.75 - 0.90)$), we obtain the nuclear matter parameters of the model enlisted in Table I. For details of the procedure involved in obtaining the parameters, one can refer to \cite{tkj08}.

We now go directly to the total energy density `$\varepsilon$' and pressure `$P$' 
for a given baryon density in terms of the dimensionless variable $Y=x/x_0$ 
which is given as:

\begin{eqnarray}
\label{ep0}
\varepsilon
&=&
 \frac{2}{\pi^2}\int^{k_B}_0 k^{2}dk{\sqrt{k^2+m_B^{\star 2}}}
         +  \frac{m_B^2(1-Y^2)^2}{8c_{\sigma B}} \nonumber \\
        &-& \frac{m_B^2 B}{12c_{\omega B}c_{\sigma B}}(1-Y^2)^3
        + \frac{m_B^2 C}{16c_{\omega B}^2c_{\sigma B}}(1-Y^2)^4 \nonumber \\
        &+& \frac{1}{2Y^2}{c_{\omega_{B}} \rho_B^2}
        +\frac{1}{2}m_{\rho}^{2}\rho_{03}^2 \nonumber \\
        &+& \frac{1}{\pi^{2}}\sum_{\lambda=e,\mu^{-}}\int^
           {k_\lambda}_0 k^{2}dk{\sqrt{k^2+m^2_{\lambda}}}\ ,
\end{eqnarray}
\begin{eqnarray}
P &=&
          \frac{2}{3\pi^2}\int^{k_B}_0\frac{k^{4}dk}
          {{\sqrt{k^2+m_B^{\star 2}}}}
        - \frac{m_B^2(1-Y^2)^2}{8c_{\sigma B}} \nonumber \\
        &+& \frac{m_B^2 B}{12c_{\omega B}c_{\sigma B}}(1-Y^2)^3
        - \frac{m_B^2 C}{16c_{\omega B}^2c_{\sigma B}}(1-Y^2)^4 \nonumber \\
        &+& \frac{1}{2Y^2}{c_{\omega_{B}} \rho_B^2}
        + \frac{1}{2}m_{\rho}^{2}\rho_{03}^2\ \nonumber \\
        &+& \frac{1}{3\pi^2}\sum_{\lambda=e,\mu^{-}}\int^{k_\lambda}_0
          \frac{k^{4}dk}{{\sqrt{k^2+m^2_{\lambda}}}}
\end{eqnarray}

The terms in eqns. (3) and (4) with the subscript $`B'$ should be interpreted 
as sum over all the states of the baryonic octets.
The meson field equations for the $\sigma$, $\omega$ and $\rho-$mesons are then solved
self-consistently at a fixed baryon density to obtain the respective field strengths.
The EOS for the $\beta-$equilibrated for the hyperon rich matter is obtained
with the requirements of conservation of total baryon number and charge
neutrality condition \cite{tkj06}. Using the computed EOS for the neutron star sequences, we calculate the structural properties of neutron stars with and without hyperon core.

\section{Results}

\begin{figure}[ht]
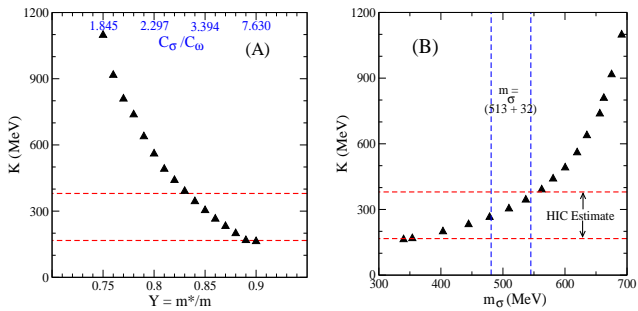

\vskip 0.3in
\begin{center}
\includegraphics[width=4cm,height=4cm,angle=0]{effm-k.eps}
\hskip 0.1in
\includegraphics[width=4cm,height=4cm,angle=0]{sigma-k.eps}
\end{center}
\caption{(A)- Nuclear matter incompressibility as a function of the nucleon effective mass for the parameters of the present model at $\rho_0 = 0.153 fm^{-3}$. Also plotted is the corresponding ratio of scalar to vector coupling on the opposite x-axis. (B)- Incompressibility as a function of obtained sigma meson mass for various parameter sets. The upper and the lower limit for incompressibility inferred from Heavy Ion Collision data \cite{data02} $K = (167-380)$ MeV is shown with horizontal lines. Recent experimental scalar meson mass limit ($m_{\sigma} = 513 \pm 32$) MeV \cite{mura02} is depicted with vertical lines}
\label{effm}
\end{figure}

Nuclear matter saturation is a consequence of the interplay between the attractive (scalar) and the repulsive (vector) forces and hence the variation in the coupling strength effects other related properties as well. Fig. \ref{effm}(A) reflects the same, where we have plotted the nuclear incompressibility for the evaluated parameter sets of the present model as a function of the nucleon effective mass. For better correlation between them, the corresponding ratio of the scalar and vector coupling is also indicated. On comparison with the incompressibility bound inferred from heavy ion collision experiment (HIC)\cite{data02}, we find that the EOS with lower nucleon effective mass is ruled out. The model favors EOS for which the nucleon effective mass $m^{\star}/m > 0.82$. It can also be seen that the EOS becomes much softer with increasing ratio of $C_{\sigma}/C_{\omega}$.

\begin{table*}
\caption{Parameter sets of the effective chiral model that satisfies the nuclear matter saturation properties such as binding energy per nucleon $B/A - m = -16.3 ~MeV$, nucleon effective mass $Y = m^{\star}/m = (0.75 - 0.90)$ and the asymmetry energy coefficient is $J \approx 32$ MeV at saturation density $\rho_0$ $=0.153 fm^{-3}$. The nucleon, the vector mesons ($\omega$ \& $\rho_0^3$) masses are taken to be 939 MeV, 783 MeV and 770 MeV respectively and $c_{\sigma} = (g_{\sigma}/ m_{\sigma})^2$, $c_{\omega} = (g_{\omega}/ m_{\omega})^2$ and $c_{\rho} = (g_{\rho}/ m_{\rho})^2$ are the corresponding coupling constants. $B = b/m^2$ and $C = c/ m^4$ are the higher order constants in the scalar field. Also given is the scalar meson mass `$m_{\sigma}$', the pion decay constant `$f_{\pi}$' and the nuclear matter incompressibility ($K$) at $\rho_0$. The maximum mass and radius of neutron star composed of nucleon only matter ($M_N, ~~R_N$) and hyperon rich matter $(M_H,~~~R_H)$ for static case are also tabulated. The last column shows the central density of the star composed of nucleon ($\varepsilon_N$) and hyperon ($\varepsilon_H$).}
\begin{center}
\begin{tabular}{cccccccccccccccc}
\hline
\hline
\multicolumn{1}{c}{set}&
\multicolumn{1}{c}{$c_{\sigma}$}&
\multicolumn{1}{c}{$c_{\omega}$} &
\multicolumn{1}{c}{$c_{\rho}$} &
\multicolumn{1}{c}{$B$} &
\multicolumn{1}{c}{$C$} &
\multicolumn{1}{c}{$m_{\sigma}$} &
\multicolumn{1}{c}{$Y$} &
\multicolumn{1}{c}{$f_{\pi}$} &
\multicolumn{1}{c}{$K$} &
\multicolumn{1}{c}{$M_N$($M_H$)} &
\multicolumn{1}{c}{$R_N~ (R_H)$} &
\multicolumn{1}{c}{$\varepsilon_{N}~(\varepsilon_{H})$} \\
\multicolumn{1}{c}{ } &
\multicolumn{1}{c}{($fm^2$)} &
\multicolumn{1}{c}{($fm^2$)} &
\multicolumn{1}{c}{($fm^2$)} &
\multicolumn{1}{c}{($fm^2$)} &
\multicolumn{1}{c}{($fm^4$)}&
\multicolumn{1}{c}{(MeV)} &
\multicolumn{1}{c}{ } &
\multicolumn{1}{c}{(MeV)} &
\multicolumn{1}{c}{($MeV$)} &
\multicolumn{1}{c}{$(M/ M_{\odot})$} &
\multicolumn{1}{c}{$(km)$} &
\multicolumn{1}{c}{$(10^{15} g cm^{3})$} \\
\hline
1   &5.916  &3.207  &5.060  &1.411   &1.328    &691.379  &0.75  &110.185  &1098 &2.49 (2.04)& 14.2 (20.8) & 1.26 (0.59) \\
2   &6.047  &3.126  &5.087  &0.822   &0.022    &675.166  &0.76  &111.601  &916 &2.47 (2.02)& 14.2 (20.7) & 1.26 (0.62) \\
3   &6.086  &3.031  &5.107  &0.485   &0.174    &662.642  &0.77  &113.346  &809 &2.44 (1.98)& 14.0 (20.5)& 1.30 (0.62) \\
4   &6.005  &2.933  &5.131  &0.582   &2.650    &656.183  &0.78  &115.238  &737 &2.41 (1.95)& 13.9 (20.4) & 1.35 (0.62) \\
5   &6.172  &2.825  &5.155 &-0.261   &0.606    &635.287  &0.79  &117.403  &638 &2.38 (1.92)& 13.7 (20.2) & 1.40 (0.65) \\
6   &6.223  &2.709  &5.178 &-0.711   &0.748    &619.585  &0.80  &119.890  &560 &2.34 (1.89)& 13.6 (20.0) & 1.44 (0.65)\\
7   &6.325  &2.585  &5.200 &-1.381   &0.089    &600.270  &0.81  &122.740  &491 &2.30 (1.85)& 13.4 (19.8) & 1.49 (0.68)\\
8   &6.405  &2.451  &5.222 &-1.990   &0.030    &580.876  &0.82  &126.039  &440 &2.26 (1.81)& 13.2 (19.5) & 1.54 (0.68)\\
{\bf 9}   & {\bf 6.474}  & {\bf 2.323}  & {\bf 5.242} &{\bf -2.533}   &{\bf 0.300}  &{\bf 562.500}  &{\bf 0.83}  &{\bf 129.465}  &{\bf 391} &{\bf 2.22 (1.77)} & {\bf 13.0 (19.3)} & {\bf 1.60 (0.71)}\\
10  &6.598  &2.159  &5.265 &-3.340   &0.445    &536.838  &0.84  &134.378  &344 &2.16 (1.72)& 12.8 (19.0) & 1.65 (0.71)\\
{\bf 11}  &{\bf 6.772}  &{\bf 1.995}  &{\bf 5.285} &{\bf -4.274}   &{\bf 0.292}  &{\bf 509.644}  &{\bf 0.85}  &{\bf 139.710}  &{\bf 303} &{\bf 2.10 (1.66)}& {\bf 12.6 (18.7)} & {\bf 1.76 (0.75)} \\
12  &7.022  &1.823  &5.305 &-5.414   &0.039    &478.498  &0.86  &146.131  &265 &2.04 (1.62)& 12.3 (18.3) & 1.82 (0.78)\\
{\bf 13}  &{\bf 7.325}  &{\bf 1.642}  &{\bf 5.324} &{\bf -6.586}   &{\bf 0.571} &{\bf 444.614}  &{\bf 0.87}  &{\bf 153.984}  &{\bf 231} &{\bf 1.97 (1.56)} & {\bf 12.0 (17.9)} & {\bf 1.94 (0.82)} \\ 
14  &7.865  &1.451  &5.343 &-8.315   &0.502  &403.303  &0.88  &163.824  &199 &1.89 (1.51)& 11.8 (17.5) & 2.08 (0.86) \\
15  &8.792  &1.249  &5.362 &-10.766  &0.354    &353.960  &0.89  &176.552  &168 &1.80 (1.44)& 11.4 (16.9) & 2.30 (0.94)\\
16  &7.942  &1.041  &5.388 &-6.908   &15.197   &339.910  &0.90  &193.437  &163 &1.71 (1.37)& 11.0 (16.3) & 2.54 (1.03)\\
\hline
\hline
\end{tabular}
\end{center}
\end{table*}

Figure \ref{effm}(B) shows the variation of incompressibility as a function of scalar meson mass obtained for various parameter sets. Recent experimental estimate for scalar meson mass  $m_{\sigma} = 513 \pm 32$ MeV \cite{mura02} is compared with the present calculation. Here we find that the EOS with $Y = (0.84 - 0.86)$ seems to agree with the combined constraint from the HIC flow data and the experimental meson mass range.

\begin{figure}
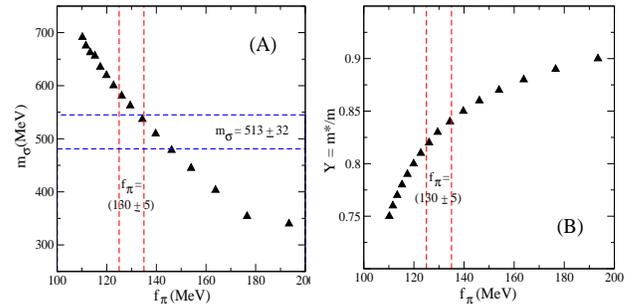

\begin{center}
\includegraphics[width=4cm,height=4cm,angle=0]{sigma-fpi.eps}
\includegraphics[width=4cm,height=4cm,angle=0]{fpi-y.eps}
\end{center}
\caption{(A)- Sigma meson mass as a function of the pion decay constant at a fixed saturation density ($\rho_0 = 0.153 fm^{-3}$). (B)- Nucleon effective mass as a function of the pion decay constant. The experimental limit of the scalar meson mass and the pion decay constant are indicated with the horizontal and the vertical lines.}
\label{fpi}
\end{figure}

Figure \ref{fpi}(A) shows the obtained sigma mass as a function of the vacuum value of the pion decay constant. The experimental bound of the pion decay constant seems to agree with slightly higher value of $m_{\sigma}$, which agree with the upper bound of the experimental bound on $m_{\sigma}$ \cite{mura02}. Figure \ref{fpi}(B) shows the nucleon effective mass $Y = m^{\star}/m$ as a function of the pion decay constant. The constraint of $f_{\pi}$ agree with EOS with $Y = (0.82 -0.84)$, however the corresponding incompressibility lies on the higher side of presently acceptable bounds (\cite{k2} - \cite{k6}).

\begin{figure}
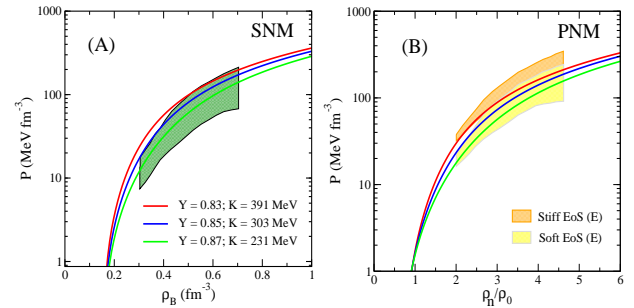

\vskip 0.3in
\begin{center}
\includegraphics[width=4cm,height=4cm,angle=0]{snm-p-expt.eps}
\includegraphics[width=4cm,height=4cm,angle=0]{expt-pnm.eps}
\end{center}
\caption{Comparison of the selected EOS with the heavy ion collision data at high densities \cite{data02}. (A)- Comparison with the Symmetric Nuclear Matter data (SNM). (B)- Comparison with the Pure Neutron Matter data (PNM).}
\label{snm}
\end{figure}

Fig. \ref{snm}(A) display the pressure as a function of baryon density up to nearly 6$\rho_0$ for the selected parameters of the model for symmetric nuclear matter. The shaded region corresponds to the experimental HIC data \cite{data02} for symmetric nuclear matter (SNM). Among the three theoretical calculations shown, the EOS with Y = 0.85 \& 0.87 agree very well with the collision data. Precisely, the third set (K = 231 MeV) completely agree with the flow data in the entire density span of $2 < \rho_B/\rho_0 < 4.6$. In Fig. \ref{snm}(B), the case of pure neutron matter (PNM) is compared with the experimental flow data. The experimental flow data is categorized in terms of stiff or soft based on whether the density dependence of the symmetry energy term is strong or week \cite{pra88}. The EOS predicted by the present model seems to rather lie on the softer regime. However, the EOS with $Y = 0.87, K = 231$ MeV though satisfy the combined constraint rather well, is not consistent with the vacuum value of the pion decay constant.

\begin{figure}
\begin{center}
\includegraphics[width=6cm,height=6cm,angle=0]{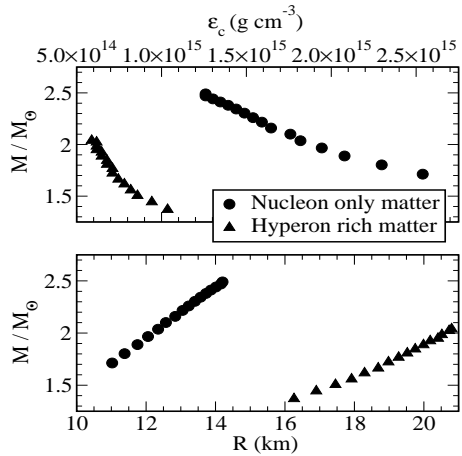}
\end{center}
\caption{({\it Upper panel:-}) The mass of the star obtained as a function of the central density of the star. ({\it Lower panel:-}) The mass of the star as a function of the radius of the star. The two cases considered here are the star composed of nucleons only ($n, p, e, \mu$) shown with the filled circles and the one composed with hyperons, shown with the triangles.}
\label{star}
\end{figure}

In Fig \ref{star}, we plot the results for the global properties of the static star, obtained from the tabulated parameters of the model for two cases. In one, the star is composed purely of $\beta-$equilibrated nucleon only matter ($n,p,e,\mu$) and in the other case, the star is composed of matter which is Hyperon rich. To study hyperon rich matter, we fixed the hyperon couplings for all the parameter sets, so as to yield the binding energy of $\Lambda^0$ at $\rho_0 = -30 MeV$. This is obtained by fixing the scalar coupling $x_{\sigma}=x_{\sigma N}/ x_{\sigma H} = 0.70$ and varying the strength of the vector counterpart to yield the binding of $\Lambda$ as suggested by hypernuclei experiments. In the upper panel of the figure, the maximum mass of the star is plotted as a function of the central density of the star. Here it can be seen that the star with nucleon matter is much massive than those composed of hyperons. However in case of star composed of nucleons, the central density ranges from ($4.5 - 9.5$)$\times ~\rho_0$, but with hyperons, the central density of the star falls in the range $\approx$~($2 - 4$)$\times ~\rho_0$ and the resulting radius increases for the later case. For all the cases given in Table I, we find nearly ($46 -48$) \% increase in the star radius, when we move from nucleonic star to hyperon rich star matter. The corresponding decrease in maximum mass is $\approx (18 - 20)$\%. This feature is very much evident from the lower panel of Fig. \ref{star}, where we have plotted the maximum mass of the star obtained as a function of the star radius. In the transition of nucleon only matter to hyperon rich neutron star matter, the compactness ratio (M/R) of the star falls from ($0.15 - 0.17$) to ($0.09 - 0.10$). 

\section{Summary}

The effective chiral model provides a natural framework to interlink the standard state properties of nuclear matter with the vacuum correlations, such as the pion decay constant. We find that the pion decay constant is experimentally well known quantity in comparison to other derived quantities such as the nuclear incompressibility and $\sigma-$meson mass, that can put stringent constraint on the model parameters. Experimentally determined effective mass from scattering of neutron over $Pb$ nuclei \cite{nuclei} seems to go well with the present model. Both of them favor higher value for nucleon effective mass. On a comparative analysis of the resulting EOS with that of the HIC data for symmetric nuclear matter as well as pure neutron matter, parameter set with $Y = 0.85; K \approx 300$ MeV seems to be the ideal parameterization of the present model. The resulting scalar meson mass $m_{\sigma} \approx 510 MeV$, is also consistent with the experimentally observed masses \cite{mura02,aitala01}. We then applied the model to study the global properties of neutron stars with and without hyperons. We find that the star composed of hyperons has a lower value of central density and results in larger radius than the star without hyperons. This in turn results in the decrease in the compactness parameter of the star. Although the resulting higher central density of the nucleonic star seems to be unphysical on account that at those higher densities, it is very unlikely that nucleons would retain its identity. Further, from the astrophysical point of view (observed neutron star masses), none of the parameters can be ruled out, however it shall be interesting to study the rotational attributes of the star with hyperons/ mixed/ quark phase in order to constrain the parameters of the model. Work is in progress in this direction \cite{tkj09}.

\end{document}